\documentclass[12pt]{article}
\usepackage{graphicx}
\usepackage{amssymb}
\usepackage{amsmath}
\usepackage{bm}
\usepackage{cite}

\newcommand{\lsim}{ \mathop{}_{\textstyle \sim}^{\textstyle <} }
\newcommand{\bear}{\begin{array}}  
\newcommand {\eear}{\end{array}}
\newcommand{\bea}{\begin{eqnarray}}   
\newcommand{\eea}{\end{eqnarray}}
\newcommand{\beq}{\begin{eqnarray}}   
\newcommand{\eeq}{\end{eqnarray}}
\newcommand{\bef}{\begin{figure}}  \newcommand 
{\eef}{\end{figure}}
\newcommand{\bec}{\begin{center}}  \newcommand 
{\eec}{\end{center}}

\begin{document}

\begin{titlepage}

\begin{flushright}
IPMU 13-0077 \\
\end{flushright}

\vskip 1.35cm
\begin{center}

{\large 
{\bf Decoupling Can Revive Minimal Supersymmetric SU(5)}
}

\vskip 1.2cm

Junji Hisano$^{a,b}$, 
Daiki Kobayashi$^a$,
Takumi Kuwahara$^a$,
and
Natsumi Nagata$^{a,c}$\\

\vskip 0.4cm

{\it $^a$Department of Physics,
Nagoya University, Nagoya 464-8602, Japan}\\
{\it $^b$Kavli Institute for the Physics and Mathematics of the Universe
 (Kavli IPMU),
University of Tokyo, Kashiwa 277-8568, Japan}\\
{\it $^c$Department of Physics, 
University of Tokyo, Tokyo 113-0033, Japan}
\date{\today}

\vskip 1.5cm

\begin{abstract} 

 We revisit proton decay via the color-triplet Higgs multiplets in the
 minimal supersymmetric grand unified model with heavy
 sfermions. Although the model has been believed to be excluded due to
 the too short lifetime of proton, we have found that it is possible to
 evade the experimental constraints on the proton decay rate if the
 supersymmetric particles have masses much heavier than the electroweak
 scale. With such heavy sfermions, the 126~GeV Higgs boson is naturally
 explained, while they do not spoil the gauge coupling unification and
 the existence of dark matter candidates. Since the resultant proton
 lifetime lies in the regions which may be reached in the future
 experiments, proton decay searches may give us a chance to verify the
 scenario as well as the supersymmetric grand unified models.

\end{abstract}

\end{center}
\end{titlepage}

%%%%%%%%%%%%%%%%%%%%%%%%%%%%%%%%%%%
\section{Introduction}
%%%%%%%%%%%%%%%%%%%%%%%%%%%%%%%%%%%

The discovery of the Higgs boson \cite{:2012gk,:2012gu} has opened the
way for physics beyond the Standard Model (SM). It is certainly a
striking hint for understanding the high-energy physics, as
elementary scalar particles may play an important role in realizing
our complicated world with apparent broken symmetries. The theories with
supersymmetry (SUSY) naturally include such scalar particles; the Higgs
boson might be a superpartner of chiral fermions, called
higgsinos. Besides, there exists a set of scalar particles for each SM
fermion, as well as adjoint fermions for the SM gauge bosons. Then,
astonishingly enough, we find that with these extra particles the gauge
coupling constants of the SM are to be unified at a certain high-energy
scale with great accuracy \cite{Dimopoulos:1981yj, Marciano:1981un,
Einhorn:1981sx, Amaldi:1991cn, Langacker:1991an}. This observation
motivates us to study the supersymmetric grand unified theories (SUSY
GUTs) \cite{Dimopoulos:1981zb, Sakai:1981gr}. 

The SUSY GUTs predict an exciting phenomenon: proton decay. 
It is induced by the exchanges of the color-triplet Higgs multiplets and the
$X$-bosons, and their effects are described in terms of the
dimension-five and -six effective operators, respectively. In the minimal SUSY
SU(5) GUT \cite{Dimopoulos:1981zb, Sakai:1981gr}, which is a simple
supersymmetric extension of the original SU(5) GUT \cite{Georgi:1974sy},
the former process yields the dominant decay modes, such as $p\to
K^+\bar{\nu}$. The lifetime of the channel is estimated as $\tau (p\to
K^+\bar{\nu})\lesssim 10^{30}$~yrs \cite{Goto:1998qg, Murayama:2001ur},
with the SUSY particles, in particular those of the third generation,
assumed to have masses of around the electroweak scale. On the other
hand, the Super-Kamiokande experiment gives stringent limits on the
channels: $\tau(p\to K^+ \bar{\nu})>4.0\times 10^{33}$~yrs
\cite{Abe:2011ts}. This contradiction makes it widely believed that
the minimal SUSY SU(5) GUT has been already excluded and, therefore,
needs some extensions in order to suppress the dimension-five proton
decay.   

As is often the case with SUSY models, the SUSY GUTs are usually
discussed within the context of the low-scale supersymmetry. 
Recently, experiments at the Large Hadron Collider (LHC) provide
limits on the SUSY models. The ATLAS and CMS Collaborations have been
searching for the SUSY particles and imposed severe constraints on their
masses, especially those of squarks and gluino \cite{Aad:2012fqa,
ATLAS:2012ona, :2012mfa}. The mass bounds have began to exceed 1~TeV and,
thus, the low-energy SUSY models are confronted with difficulties. 
Moreover, the observed mass of the Higgs boson
around 126~GeV \cite{:2012gk,:2012gu} might also indicate the SUSY scale
is considerably higher than the electroweak scale; in the minimal
SUSY Standard Model (MSSM), the mass of the lightest Higgs boson is
below the $Z$-boson mass at tree level, so sufficient mass difference
between stops and top quark is required in order to raise the Higgs boson
mass through the radiative corrections \cite{Okada:1990vk, Okada:1990gg,
Haber:1990aw, Ellis:1990nz, Ellis:1991zd}.

In fact, the SUSY models with heavy SUSY particles have a lot of
attractive features \cite{Wells:2003tf, ArkaniHamed:2004fb,
Giudice:2004tc, ArkaniHamed:2004yi, Wells:2004di, Hall:2009nd,
Hall:2011jd}. First of all, since the flavor
changing neutral current (FCNC) processes and/or the electric dipole
moments induced by SUSY particles are suppressed by their masses, the
SUSY flavor and CP problems \cite{Gabbiani:1996hi} are relaxed when the
masses are considerably heavy. As mentioned to above, the heavy
sfermions yield sufficient radiative corrections to lift the Higgs mass
up to 126~GeV \cite{Giudice:2011cg, Ibe:2011aa, Ibe:2012hu,
Ibanez:2013gf}. They do not spoil the gauge coupling unification
since the sfermions form complete SU(5) multiplets. 
Actually, it turns out that the unification is improved in the
sense that the required threshold corrections at the GUT scale tend to
be reduced \cite{Hisano:2013cqa}.
As for the 
cosmology, the gravitino problem is avoided because of the high-scale
SUSY breaking, and the thermal leptogenesis scenario well works with high
reheating temperature \cite{Fukugita:1986hr}. Further, this high-scale
SUSY scenario naturally accommodates dark matter (DM) candidates, which
might be detected in future dark matter experiments directly
\cite{Hisano:2010fy, Hisano:2010ct, Hisano:2011cs, Hisano:2012wm} and
indirectly \cite{Hisano:2003ec, Hisano:2004ds, Ibe:2012hu}. Thus, with
the recent LHC results considered, the high-scale SUSY scenario is even
promising from a phenomenological point of view.

Interestingly, this scenario also provides an alternative solution to the
problem regarding the dimension-five proton decay in the minimal SUSY
GUT. The dimension-five operators generated via the color-triplet Higgs
exchange contain squarks and/or sleptons in their external lines. These
fields are to be integrated out below the SUSY scale through the wino or
higgsino exchanging processes, and then the four-Fermi operators,
suppressed by the sfermion masses, are induced. Hence, their effects are
expected to be extremely reduced when the SUSY scale is much higher than
the electroweak scale. 

In this paper, we study such possibilities within the context of the
high-scale SUSY scenario. We will find that the minimal SUSY SU(5) GUT
actually evades the constraints from the proton decay experiments with
the SUSY braking scales which naturally explain the 126~GeV Higgs boson
and the existence of dark matter in the Universe. The resultant proton
lifetime lies in the regions which may be reached in the future proton
decay experiments. Therefore, although the high-scale SUSY scenario is
hard to be probed in the collider experiments, the proton decay searches
may give us a chance to verify the scenario as well as the existence of
supersymmetry and the grand unification. 

This paper is organized as follows: In Sec.~\ref{model}, a high-scale
SUSY model which we discuss in this work and its phenomenology are
briefly explained. In the next section, we give a set of formulae for
evaluating the proton decay rate via the dimension-five operators. Then,
we show the resultant proton lifetime in the model, and compare it with
current experimental limits in
Sec.~\ref{results}. Section~\ref{conclusion} is devoted to conclusions
and discussion.

%%%%%%%%%%%%%%%%%%%%%%%%%%%%%%%%%%%
\section{High-scale SUSY}
\label{model}
%%%%%%%%%%%%%%%%%%%%%%%%%%%%%%%

To begin with, we describe a high-scale SUSY model which we deal with in
the following discussion. Assume that there exists a SUSY breaking hidden
sector where the SUSY breaking is triggered by a chiral superfield $Z$
which is not a gauge singlet. Then, with a generic form of K\"{a}hler
potential, all the scalar bosons except the lightest Higgs boson acquire
masses of 
\begin{equation}
 M_S\sim\frac{F_Z}{M_*}~,
\end{equation}
with $F_Z$ and $M_*$ the $F$-component
vacuum expectation value (VEV) of the field $Z$ and the mediation scale
of the SUSY breaking, respectively.The gravitino mass is
$m_{3/2}=F_Z/\sqrt{3}M_{\rm Pl}$, and it is the same order as the scalar
masses when $M_*$ is around the Planck scale, $M_{\rm Pl}$. The soft masses for the two doublet Higgses and the $\mu$-term are fine-tuned in order to realize the electroweak symmetry breaking at the proper scale. The gaugino
masses, on the other hand, are not generated by the dimension-five
operators like
\begin{equation}
 \int d^2\theta ~Z{\rm Tr}[W^{\alpha}W_\alpha]~,
\end{equation}
since the symmetry under which the superfield $Z$ is charged
prohibits such an operator. Here,
$W_\alpha$ denotes the gauge field-strength chiral superfield. Instead,
they are induced by the anomaly mediation mechanism
\cite{Randall:1998uk, Giudice:1998xp}:
\begin{equation}
 M_a=\frac{b_ag^2_a}{16\pi^2}m_{3/2}~,
\end{equation}
where $M_1$, $M_2$, and $M_3$ are the masses of bino, wino, and gluino,
respectively, and $b_a$ are the one-loop beta-function coefficients of
the gauge coupling constants $g_a$ ($a=1,~2,$ and 3 for U(1)$_Y$,
SU(2)$_L$, and SU(3)$_C$, respectively). This expression tells us that
the gaugino masses are suppressed by one-loop factors compared with
the gravitino mass. Similarly, since we assume that the superfield which breaks supersymmetry is not a gauge-singlet, the A-terms are generated with the suppression by the loop factors. Finally, the higgsino mass, $\mu_H$, is somewhat model-dependent; it might lie around the same order of gaugino masses when some
additional symmetries exist, or be as large as gravitino masses if 
it is generated by the K\"{a}hler potential. Thus we regard it as a free
parameter in the following discussion.

In this scenario, the lightest SUSY particle (LSP) is either
wino, which turns out to be the lightest gaugino, or the lighter
higgsino. Then, it is found that in any case the LSP may explain the
dark matter in the Universe. Indeed, the thermal relic abundance of wino
and higgsino DM with a mass of 2.7--3.0~TeV \cite{Hisano:2006nn} and 1~TeV
\cite{Cirelli:2007xd}, respectively, accounts for the observed density
of DM. With relatively small masses, the non-thermal
production of them also might be consistent with the observation
\cite{Gherghetta:1999sw, Moroi:1999zb}. The wino or higgsino DM in
this model implies sfermion masses lie around $10^2$--$10^3$~TeV.

From now on, we assume the sfermions are nearly degenerate in mass, and
their masses are collectively denoted by $M_S$. The mass is supposed to
be $M_S\simeq 10^2$--$10^3$~TeV, and either wino or higgsino is assumed
to be the LSP. Models with such a mass spectrum and their
phenomenology have been enthusiastically studied in the previous
literature \cite{Jeong:2011sg, Saito:2012bb, Sato:2012xf,
Bhattacherjee:2012ed, Hall:2012zp, ArkaniHamed:2012gw,
Moroi:2013sfa,McKeen:2013dma}.  

As mentioned to in the Introduction, the mass spectrum does not spoil
the gauge coupling unification \cite{ArkaniHamed:2004fb,
Giudice:2004tc}, since the sfermions are embedded in the complete
multiplets of SU(5). 
In fact, the unification may be improved \cite{Hisano:2013cqa}; a
renormalization group analysis reveals that all of the GUT scale
particles, especially the color-triplet Higgs multiplets, possibly lie around
the GUT scale $\simeq 10^{16}$~GeV, contrary to the case of the
low-energy SUSY. It indicates that the threshold corrections to the
gauge coupling constants at the GUT scale are reduced in the high-scale
SUSY scenario. Thus, the SUSY GUTs are still well-motivated.

%%%%%%%%%%%%%%%%%%%%%%%%%%%%%%%%
\section{Proton Decay in the Minimal SUSY SU(5) GUT}
\label{dim5}
%%%%%%%%%%%%%%%%%%%%%%%%%%%%%%%%%

In this section, we review the proton decay in the minimal
SUSY SU(5) GUT. In this model, the MSSM
matter fields are embedded in a $\bar{\bf 5}\oplus {\bf 10}$
representation. The SU(2)$_L$ singlet down-type quarks $\bar{D}$
and doublet leptons $L$ are incorporated into the
$\bar{\bf 5}$ fields, $\Phi$, while the SU(2)$_L$ singlet up-type
quarks, $\bar{U}$, doublet quarks, $Q$, and singlet leptons, $\bar{E}$,
are formed into the ${\bf 10}$ representations, $\Psi$. Here all the
superfields are expressed in terms of the left-handed chiral superfields.
The explicit form of the multiplets is 
\begin{align}
 \Phi &=
\begin{pmatrix}
 \bar{D}_1 \\
 \bar{D}_2 \\
 \bar{D}_3 \\
 E \\
 -N
\end{pmatrix}~, ~~~~~~
\Psi=\frac{1}{\sqrt{2}}
\begin{pmatrix}
 0&\bar{U}_3&-\bar{U}_2&U^{1}&D^{1} \\
 -\bar{U}_3&0&\bar{U}_1&U^{2}&D^{2}\\
 \bar{U}_2&-\bar{U}_1&0&U^{3}&D^{3}\\
 -U^{1}&-U^{2}&-U^{3}&0&\bar{E} \\
 -D^{1}&-D^{2}&-D^{3}&-\bar{E} &0
\end{pmatrix}~,
\end{align}
with
\begin{equation}
 L=
\begin{pmatrix}
 N \\ E 
\end{pmatrix}~, ~~~~~~
Q^{\alpha}=
\begin{pmatrix}
 U^{\alpha} \\ D^{\alpha}
\end{pmatrix}~.
\end{equation}
Here, $\alpha=1,2,3$ denotes the color index.
The MSSM Higgs superfields, on the other hand, are embedded
into a pair of {\bf 5} and $\bar{\bf 5}$ superfields accompanied with
the new Higgs superfields $H^\alpha_C$ and $\bar{H}_{C\alpha}$ called
the color-triplet Higgs multiplets: 
\begin{equation}
 H=
\begin{pmatrix}
 H^1_C \\ H^2_C \\ H^3_C \\ H^+_u \\ H^0_u
\end{pmatrix}
,~~~~~~\bar{H}=
\begin{pmatrix}
 \bar{H}_{C1}\\
 \bar{H}_{C2}\\
 \bar{H}_{C3}\\
 H^-_d \\ -H^0_d
\end{pmatrix}
~,
\end{equation}
where the last two components are corresponding to the MSSM Higgs superfields,
\begin{equation}
 H_u=
\begin{pmatrix}
 H^+_u \\ H^0_u
\end{pmatrix}
,~~~~~~H_d=
\begin{pmatrix}
 H^0_d \\ H^-_d
\end{pmatrix}
~.
\end{equation}

Exchanges of the color-triplet Higgs multiplets induce the
baryon-number violating interactions. They are coupled with the
ordinary matter fields by the Yukawa coupling terms in the
superpotential\footnote{ In this paper we evaluate the proton
    decay rate with the SU(5) symmetric Yukawa couplings which are
    evaluated with up-type and down-type quark masses and the
    Cabibbo-Kobayashi-Maskawa matrix, though the ratios of charged
    lepton and down-type quark masses are not necessarily consistent
    with them. We have checked that, even if the lepton masses are used for $f_{d_i}$, 
    our consequence presented below is not changed significantly. 
}:

\begin{align}
 W_{\rm Yukawa} &=
\frac{1}{4}h^{ij}\epsilon_{abcde}\Psi_i^{ab} \Psi_j^{cd}H^e -\sqrt{2}
f^{ij}\Psi_i^{ab} \Phi_{ja}\bar{H}_b~,
\label{superpotential}
\end{align}
where $i,j~=1,2,3$ and $a,b,c,\dots~=1$--$5$ represent the generations and the
SU(5) indices, respectively.
The Yukawa couplings $h^{ij}$ and $f^{ij}$ in Eq.~(\ref{superpotential})
have redundant degrees of freedom, most of which are eliminated by the
field re-definition of $\Psi$ and $\Phi$ \cite{Ellis:1978xg}. We
parametrize the couplings according to Ref.~\cite{Hisano:1992jj} as
\begin{align}
 h^{ij}&= f_{u_i}e^{i\varphi_i}\delta_{ij}~, \nonumber \\
 f^{ij}&=V^*_{ij}f_{d_j}~,
\label{yukawadef}
\end{align}
with $V_{ij}$ the Cabibbo-Kobayashi-Maskawa (CKM) matrix. The phase
factors $\varphi_i$ are subject to a constraint
\begin{equation}
 \varphi_1+\varphi_2+\varphi_3=0~,
\label{constraint}
\end{equation}
and thus two of them are independent parameters. 
The Yukawa coupling terms in Eq.~\eqref{superpotential} are written with the
component fields as follows:
\begin{align}
 W_{\rm Yukawa}&=f_{u_i}e^{i\varphi_i}\epsilon_{rs}
\overline{U}_{i\alpha}Q^{\alpha r}_i
 H_u^s -V^*_{ij}f_{d_j}\epsilon_{rs}Q^{\alpha r}_i\overline{D}_{j\alpha
} H_d^s -f_{d_j}\epsilon_{rs}V^*_{ij}\overline{E}_iL^r_j H^s_d\nonumber \\
&-\frac{1}{2}f_{u_i}e^{i\varphi_i}\epsilon_{\alpha\beta\gamma}\epsilon_{rs}
Q^{\alpha r}_i Q^{\beta s}_i H^\gamma_C
+V_{ij}^*f_{d_j}\epsilon_{rs}Q^{\alpha r}_iL^s_j\overline{H}_{C\alpha}
\nonumber \\
&+f_{u_i}e^{i\varphi_i}\overline{U}_{i\alpha}\overline{E}_iH^\alpha_C
-V^*_{ij}f_{d_j}\epsilon^{\alpha\beta\gamma}
\overline{U}_{i\alpha}\overline{D}_{j\beta}\overline{H}_{C\gamma}~.
\end{align}
Here, $r, s$ are the SU(2)$_L$ indices. In the following discussion, we
also use the flavor basis for the matter fields. The relations between the
flavor and gauge eigenstates are given as
\begin{align}
 Q_i&=
\begin{pmatrix}
 U_i^\prime \\ V_{ij}D_j^\prime
\end{pmatrix},~~~~~~L_i =
\begin{pmatrix}
 N_i^\prime \\ E_i^\prime
\end{pmatrix},\nonumber
\end{align}
\begin{equation}
 \bar{U}_i =e^{-i\varphi_i}\bar{U}_i^\prime,~~~~~~
  \bar{D}_i =  \bar{D}_i^\prime,~~~~~~
\bar{E}_i=V_{ij}\bar{E}_j^\prime~,
\end{equation}
where primes represent the flavor eigenstates.

%%%%%%%FIGURE%%%%%%%%%%%%%%%%%%%%%%
\begin{figure}[t]
\begin{center}
\includegraphics[height=50mm]{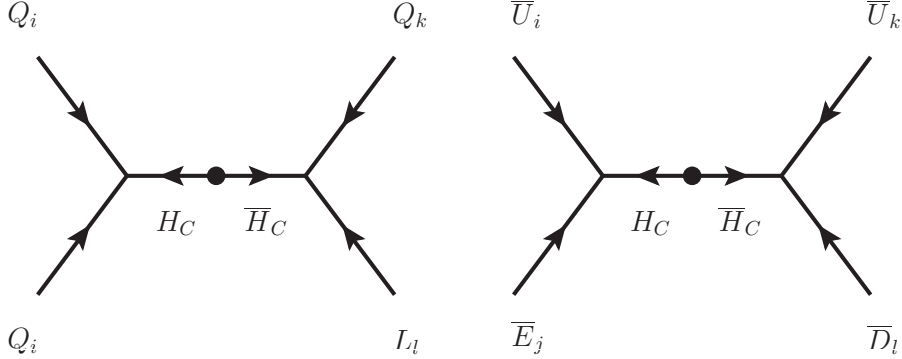}
\caption{Supergraphs which illustrate color-triplet Higgs exchanging
  processes where dimension-five effective operators for proton decay
  are induced. Bullets indicate color-triplet Higgs mass term.}
\label{fig:dim5}
\end{center}
\end{figure}
%%%%%%%%%%%%%%%%%%%%%%%%%%%%%%%%%
The Yukawa interactions of color-triplet Higgs multiplets give rise to the dimension-five baryon-number
violating operators. The processes in which the operators are induced
are illustrated by the diagrams in Fig.~\ref{fig:dim5}. After the color-triplet Higgs multiplets are decoupled, we obtain the effective
superpotential as
\begin{align}
 W_5&=+\frac{1}{2M_{H_C}}f_{u_i}f_{d_l}V^*_{kl}
e^{i\varphi_i}\epsilon_{\alpha\beta\gamma}
\epsilon_{rs}\epsilon_{tu}Q^{\alpha r}_iQ^{\beta s}_iQ^{\gamma
t}_k L^u_l\nonumber \\
&+\frac{1}{M_{H_C}}f_{u_i}e^{i\varphi_i}f_{d_l}V^*_{kl}
\epsilon^{\alpha\beta\gamma} \overline{U}_{i\alpha}\overline{E}_i
\overline{U}_{k\beta}\overline{D}_{l\gamma}~,
\label{W5}
\end{align}
which yields the dimension-five effective operators,
\begin{equation}
{\cal L}_5= \int ~ d^2\theta W_5~~+~~{\rm h.c.}\,.
\end{equation}

%%%%%%%FIGURE%%%%%%%%%%%%%%%%%%%%%%
\begin{figure}[t]
\begin{center}
\includegraphics[height=60mm]{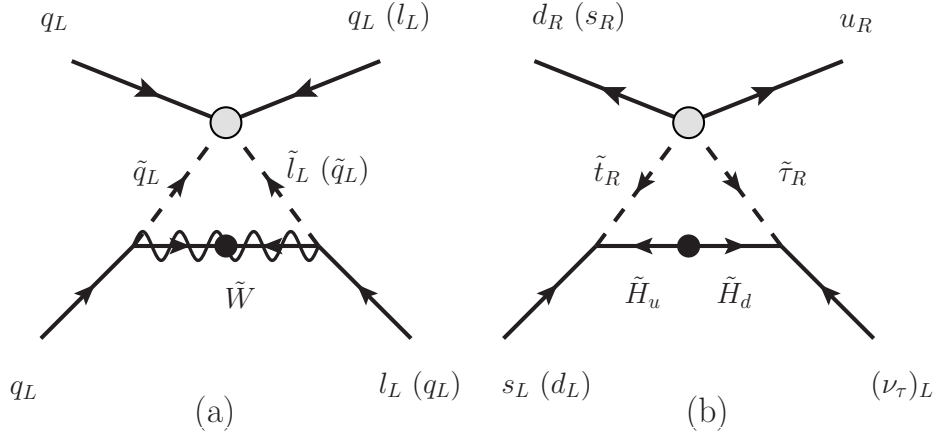}
\caption{One-loop diagrams which yield the baryon-number violating
 four-Fermi operators. Diagrams (a) and (b) are generated by charged wino
 and higgsino exchanging processes, respectively. Gray dots indicate 
 dimension-five effective interactions, while black dots represent 
  wino or higgsino mass terms.}
\label{fig:1loop}
\end{center}
\end{figure}
%%%%%%%%%%%%%%%%%%%%%%%%%%%%%%%%%

The effective operators contain sfermions in their external lines. Below
the SUSY breaking scale, $M_S$,  these sfermions turn into the SM fermions via
the charged wino and higgsino exchanging processes shown in
Fig.~\ref{fig:1loop}. In this figure, the gray and black dots indicate
the dimension-five effective interactions and the mass terms for wino or
higgsino, respectively. The first operator in Eq.~\eqref{W5} contributes
to the diagram (a), while the second one induces the diagram
(b). Although the contribution of the diagram (b) is suppressed by the
CKM matrix elements as it is generated in the flavor changing
process, it is found to be sizable because of the large Yukawa couplings
of the third generation fermions \cite{Goto:1998qg, Lucas:1996bc}. 
The contributions of flavor-conserving neutral gauginos and higgsino exchange are in general suppressed by the Yukawa couplings of 
the first generation, thus negligible. Among them, the gluino contribution might
be sizable because of the large coupling. It turns out, however, that
the gluino contribution vanishes in the limit where squarks are
degenerate in mass, and we consider such a case in the following
calculation. When sfermion mass matrices have large flavor mixing, the
contributions may also be significant, though we do not take into
account such a situation for simplicity.

After the
electroweak symmetry breaking, the charged wino and higgsino are mixed
with each other. In the following calculation, however, we neglect the
effect since we mainly consider the case where $M_2, \mu_H \gg m_W$ with
$m_W$ the mass of $W$-boson. When the masses of wino and higgsino are
nearly degenerate, the mixing effects might be significant. It is
straight-forward to modify the formulae obtained below in such a case.

Evolving the four-Fermi operators from the SUSY breaking scale to the
hadron scale ($\sim1$~GeV) according to the renormalization group
equations (RGEs), we finally obtain the effective operators for proton
decay as
\begin{align}
 {\cal L}_6=\frac{\alpha_2^2}{M_{H_C}m_W^2\sin 2\beta}\biggl[
&2F(M_2, M_S^2)
\sum_{i,j=2,3}\overline{m}_{u_i}\overline{m}_{d_j}V_{u_i d}V_{u_i s}V^*_{ud_j}e^{i\varphi_{i}}
\nonumber \\
&\times A_{R}^{(i,j)}\epsilon_{\alpha\beta\gamma}\bigl\{
(u^\alpha_L d^\beta_L)(\nu_{Lj}s_L^\gamma )
+(u^\alpha_L s^\beta_L)(\nu_{Lj}d_L^\gamma )
\bigr\}\nonumber \\
-\frac{\overline{m}_t^2\overline{m}_\tau V^*_{tb}e^{i\varphi_1}}
{m_W^2\sin 2\beta}
&F(\mu_H, M_S^2)
\overline{A}_R\epsilon_{\alpha\beta\gamma} 
\bigl\{
\overline{m}_dV_{ud}V_{ts}(u_R^\alpha d_R^\beta)(\nu_\tau s_L^\gamma)
+\overline{m}_sV_{us}V_{td}(u_R^\alpha s_R^\beta)(\nu_\tau d_L^\gamma)
\bigr\}\biggr]\nonumber \\
&~+{\rm h.c.},
\label{L6}
\end{align}
where we use the two-component spinor notation for the SM fermion
fields; all of the quarks are written in the flavor basis though primes
are omitted for simplicity;
$M_S$ is the mass of sfermions with all the sfermions assumed to be
degenerate in mass;
$\overline{m}_q$ are the masses of quarks defined in the $\overline{\rm
DR}$ scheme at the scale of $\mu =2$~GeV; $u_2$, $d_2$, $u_3$, and $d_3$
denote $c$, $s$, $t$, and $b$ quarks, respectively; $\alpha_2\equiv
g_2^2/ 4\pi$ with $g_2$ the SU(2)$_L$ gauge coupling constant at the
electroweak scale;
$\tan \beta \equiv \langle H_u^0\rangle / \langle H_d^0\rangle$; 
$A_R^{(i,j)}$ and $\overline{A}_R$ in Eq.~\eqref{L6}
represent the renormalization factors. These factors include the
renormalization effects for both the couplings and the effective
operators. The estimation of these factors is carried out in
Appendix~\ref{Renormalization}. 

The loop function $F(M, M_S^2)$ is
given as 
\begin{equation}
 F(M, M_S^2)=M\biggl[
\frac{1}{M_S^2-M^2}-\frac{M^2}{(M^2_S-M^2)^2}\ln\biggl(
\frac{M_S^2}{M^2}
\biggr)
\biggr]~,
\end{equation}
where $M$ is either the wino mass, $M_2$, or the higgsino mass,
$\mu_H$. In the limit of $M\ll M_S$, the function leads to
\begin{equation}
 F(M, M_S^2)\to \frac{M}{M_S^2},~~~~~(M_S\gg M)~,
\end{equation}
while in the limit of $M\to M_S$, it follows that
\begin{equation}
 F(M, M_S^2)\to \frac{1}{2M_S},~~~~~(M\to M_S)~.
\end{equation}
Note that the function is proportional to $M$,  when $M\lsim M_S$. For
this reason, the contribution of the diagrams in Fig.~\ref{fig:1loop}
is enhanced when the masses of the exchanged particles are large. In
particular, in the case of $\mu_H \gg M_2$, the higgsino exchange
contribution (the diagram (b) in Fig.~\ref{fig:1loop}) dominates the
wino exchange one.  We also find from the behavior of the loop function that the transition amplitude is
considerably suppressed when the sfermions have sufficiently large
masses. Thus, we expect that the experimental constraints on the proton
decay rate may be avoided in the high-scale SUSY scenario.

The effective operators in Eq.~\eqref{L6} are written in terms of
partons. In order to derive the decay amplitude for proton, we need
to obtain the matrix elements of the quarks appearing in the
operators between the proton state $\vert p\rangle$ and the kaon state
$\vert K^+\rangle$. With the matrix elements, we finally obtain the partial
decay widths, $\Gamma (p \to K^+ \bar{\nu}_\mu)$ and $\Gamma (p \to K^+
\bar{\nu}_\tau)$, and the sum of them well approximates the total decay
rate of the $p\to K^+ \bar{\nu}$ channel. Explicit formulae for the
decay widths as well as the determination of the matrix elements are
presented in Appendix.~\ref{Formulas}.

Before concluding this section, we briefly comment on the proton decay
via the SU(5) gauge boson exchange. The SU(5) gauge bosons, called
$X$-bosons, give rise to the dimension-six operators which contribute to
proton decay by the $p\to \pi^0 e^+ $ channel. 
In Ref.~\cite{Hisano:2013cqa}, it is pointed out that the GUT scale in
the high-scale SUSY tends to be slightly lower than that in the
low-energy SUSY. Thus, the proton decay rate in this channel is expected to
be enhanced. It turns out, however, that the resultant lifetime is
generally long enough \cite{Hisano:2012wq} to evade the current
experimental bound, $\tau(p\to \pi^0 e^+)> 1.29\times 10^{34}$~yrs
\cite{:2012rv}. Thus, we ignore the contribution in the following
calculation.

%%%%%%%%%%%%%%%%%%%%%%%%%%%%%%%%%%
\section{Results}
\label{results}
%%%%%%%%%%%%%%%%%%%%%%%%%%%%%%%%%%

Now we show numerical results of the proton decay lifetime in the high-scale
SUSY scenario. We will see below that the resultant lifetime is well
above the current experimental limits in a wide range of parameter
region. 

First, we consider the case where the higgsino mass is of the order of
the sfermion masses, $M_S$. In this case, the higgsino exchange contribution (the
diagram (b) in Fig.~\ref{fig:1loop}) dominates the wino exchange one,
as mentioned to in the previous section. For this reason, the lifetime 
has little dependence on the additional phases, $\varphi_{i}$ in
Eq.~\eqref{yukawadef}, as well as the wino mass. Thus, it is possible to
make a robust prediction for the proton decay lifetime. As the
right-handed stop and stau run in the loop in the higgsino exchanging
diagram, $M_S$ should be regarded as their masses, which we assume to be
degenerate for brevity. For $M_S=\mu_H$, the proton
lifetime $\tau_p$ is approximately given as\footnote
{The renormalization factor reduces the proton decay rate for larger $M_S$
when $M_S=\mu_H=(10^2$--$10^5$)~TeV, as described in
Fig.~\ref{fig:arbar} in Appendix.~\ref{Renormalization}.}
\begin{equation}
\tau_p
\simeq 4\times 10^{35}\times\sin^4 2\beta~
\biggl(\frac{0.1}{\overline{A}_R}\biggr)^2
 \biggl(\frac{M_S}{10^2~{\rm TeV}}\biggr)^2
 \biggl(\frac{M_{H_C}}{10^{16}~{\rm GeV}}\biggr)^2
~~{\rm yrs}
~,
\label{approx_lifetime}
\end{equation}
and found to be well above the current experimental limits, 
$\tau(p\to K^+ \bar{\nu})>4.0\times 10^{33}$~yrs
\cite{Abe:2011ts}, with the SUSY scale being much higher than the
electroweak scale. 

Let us investigate it in detail. To begin with, we consider the mass of
the color-triplet Higgs multiplets, $M_{H_C}$. 
As mentioned to in Sec.~\ref{model}, through the RGE analysis discussed
in Refs.~\cite{Hisano:1992mh, Hisano:1992jj} with requiring the gauge
coupling unification, one finds that $M_{H_C}$
may be around the GUT scale in the case of high-scale SUSY
\cite{Hisano:2013cqa}. The prediction is, however, quite sensitive to
the mass spectrum below the GUT scale, especially to the masses of
higgsinos and gauginos. Thus,
in the following discussion, we just fix
$M_{H_C}$ to be around the GUT scale. One easily obtains proton
lifetimes corresponding to other values of $M_{H_C}$ by using the
power law given in Eq.~\eqref{approx_lifetime}.

%%%%%%%FIGURE%%%%%%%%%%%%%%%%%%%%%%
\begin{figure}[t]
\begin{center}
\includegraphics[height=75mm]{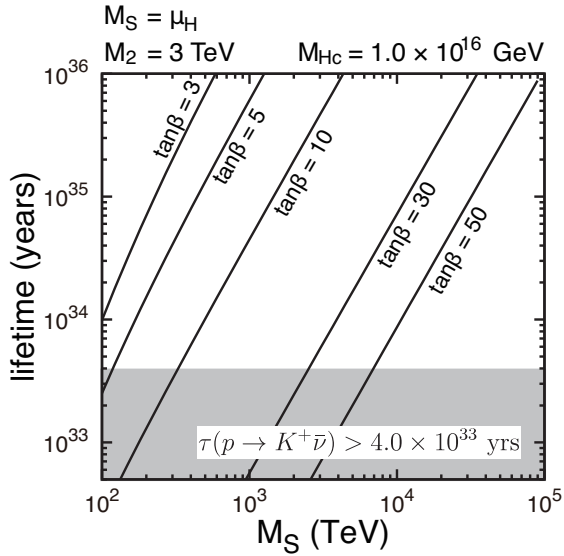}
\caption{Lifetime of $p\to K^+ \bar{\nu}$ mode as functions of
 $M_S=\mu_H$. Wino mass is set to be 3~TeV and $M_{H_C}=1.0\times
 10^{16}~{\rm GeV}$. Solid lines correspond to 
 $\tan \beta=3,5,10,30$, and 50 from left-top to right-bottom,
 respectively.  Shaded region is excluded by the current experimental bound,
$\tau(p\to K^+ \bar{\nu})>4.0\times 10^{33}$~yrs \cite{Abe:2011ts}.}
\label{fig:hig16}
\end{center}
\end{figure}
%%%%%%%%%%%%%%%%%%%%%%%%%%%%%%%%%

%%%%%%%FIGURE%%%%%%%%%%%%%%%%%%%%%%
\begin{figure}[t]
\begin{center}
\includegraphics[height=75mm]{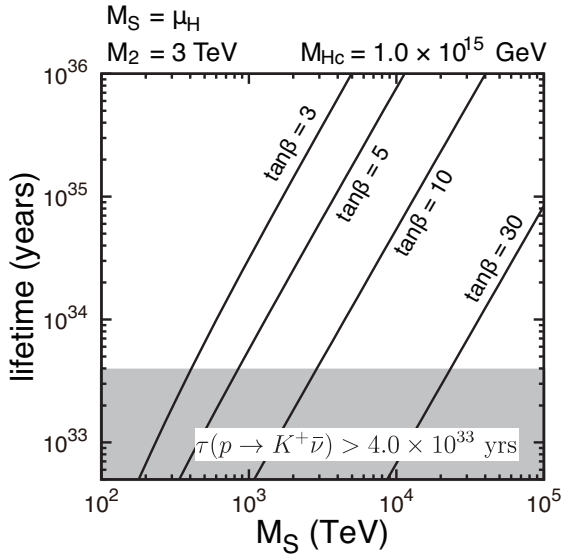}
\caption{Lifetime of $p\to K^+ \bar{\nu}$ mode as functions of
 $M_S=\mu_H$. Wino mass is set to be 3~TeV and $M_{H_C}=1.0\times
 10^{15}~{\rm GeV}$. Solid lines correspond to 
 $\tan \beta=3,5,10$, and 30 from left-top to right-bottom,
 respectively. Shaded region is excluded by the current experimental bound,
 $\tau(p\to K^+ \bar{\nu})>4.0\times 10^{33}$~yrs \cite{Abe:2011ts}.}
\label{fig:hig15}
\end{center}
\end{figure}
%%%%%%%%%%%%%%%%%%%%%%%%%%%%%%%%%

In Fig.~\ref{fig:hig16}, we present the lifetime of the $p\to K^+
\bar{\nu}$ mode as functions of $M_S=\mu_H$. Here, the wino mass is set to
be 3~TeV, while the result scarcely depends on the mass as long as
$M_2\ll M_S$. The color-triplet Higgs mass is fixed to $M_{H_C}=1.0\times
10^{16}~{\rm GeV}$. The solid lines are for $\tan \beta=3,5,10,30$, and 50
from left-top to right-bottom, respectively. The shaded region is
excluded by
the current experimental bound, $\tau(p\to K^+ \bar{\nu})>4.0\times
10^{33}$~yrs \cite{Abe:2011ts}. The figure illustrates the
behavior presented in Eq.~\eqref{approx_lifetime}. Moreover, it is found
that the proton decay lifetime in the high-scale SUSY scenario may evade
the experimental constraints, especially for small $\tan \beta$ and high
SUSY breaking scales. We also show a similar plot for a relatively small
value of $M_{H_C}$ in Fig.~\ref{fig:hig15}, where the mass is taken to
be $1.0\times 10^{15}$~GeV. The wino mass is again set to be
$M_2=3$~TeV, and the solid lines correspond to $\tan \beta=3,5,10$,
and 30 from left-top to right-bottom, respectively. We see that the
relation between the results presented in Figs.~\ref{fig:hig16} and
\ref{fig:hig15} is well explained by the simple power law in
Eq.~\eqref{approx_lifetime}, though the renormalization factors may also
be changed with different values of $M_{H_C}$. For this reason, we just
fix $M_{H_C}=1.0\times 10^{16}$~GeV in the following analysis. One
easily read other results with different values of $M_{H_C}$ by using
the relation given in Eq.~\eqref{approx_lifetime}.

%%%%%%%FIGURE%%%%%%%%%%%%%%%%%%%%%%
\begin{figure}[t]
\begin{center}
\includegraphics[height=75mm]{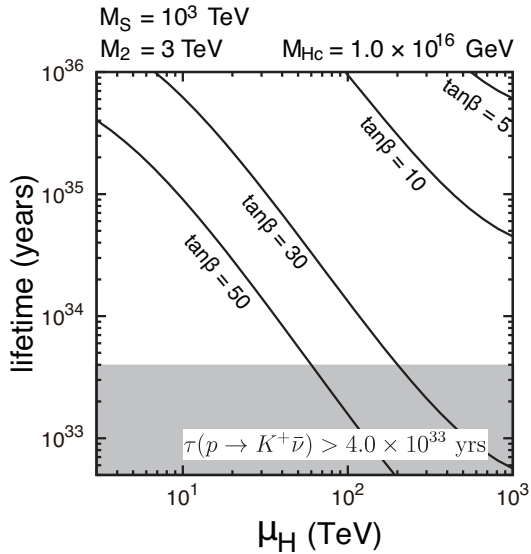}
\caption{Lifetime of $p\to K^+ \bar{\nu}$ mode as functions of
 $\mu_H$. Wino, sfermion and color-triplet Higgs masses are set to be
 $M_2=3$~TeV, $M_S=10^3$~TeV, and $M_{H_C}=1.0\times
 10^{16}~{\rm GeV}$, respectively. Solid lines correspond to 
 $\tan \beta=5,10,30$, and 50 from right-top to left-bottom,
 respectively. Shaded region is excluded by the current experimental bound,
 $\tau(p\to K^+ \bar{\nu})>4.0\times 10^{33}$~yrs \cite{Abe:2011ts}.}
\label{fig:win3tev}
\end{center}
\end{figure}
%%%%%%%%%%%%%%%%%%%%%%%%%%%%%%%%%

Next, we consider the case where the higgsinos are lighter than the
sfermions. In this case, the lifetime depends on the new phases
appearing in Eq.~\eqref{yukawadef}. Here, we take the phases so that
they yield the maximal amplitude for the proton decay rate, {\it i.e.},
we require that each term in Eqs.~\eqref{cmu} and \eqref{ctau} be
constructive. This requirement together with the constraint
\eqref{constraint} uniquely determines all of the phases
$\varphi_i$. Since the choice of phases gives the maximal proton decay
rate, we are to obtain the most stringent limit on the parameters. In
addition, we assume that both the higgsino and wino mass parameters are
real and positive. However, as long as one chooses the phases
constructively, the results would not change since it is possible to
include the extra phases of the higgsino and wino masses into the
redefinition of the phases $\varphi_i$.

%%%%%%%FIGURE%%%%%%%%%%%%%%%%%%%%%%
\begin{figure}[t]
\begin{center}
\includegraphics[height=75mm]{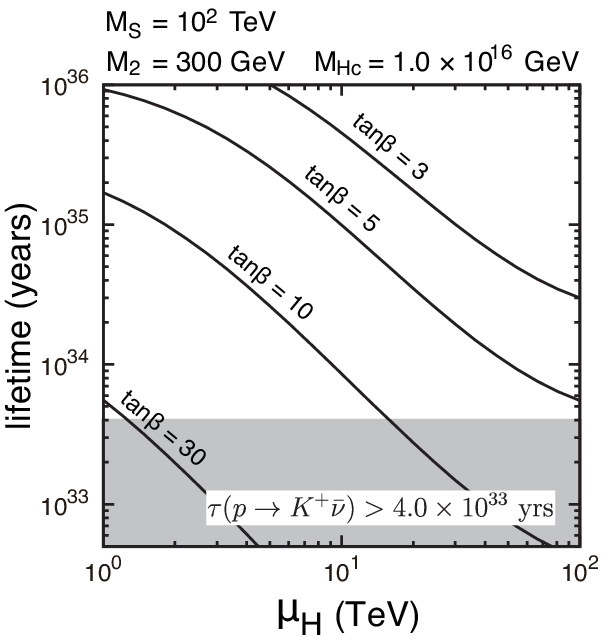}
\caption{Lifetime of $p\to K^+ \bar{\nu}$ mode as functions of
 $\mu_H$. Wino, sfermion, and color-triplet Higgs masses are set to be
 $M_2=300$~GeV, $M_S=10^2$~TeV, and $M_{H_C}=1.0\times
 10^{16}~{\rm GeV}$, respectively. Solid lines correspond to 
 $\tan \beta=3,5,10$, and 30 from right-top to left-bottom,
 respectively. Shaded region is excluded by the current experimental bound,
 $\tau(p\to K^+ \bar{\nu})>4.0\times 10^{33}$~yrs \cite{Abe:2011ts}.}
\label{fig:win300gev}
\end{center}
\end{figure}
%%%%%%%%%%%%%%%%%%%%%%%%%%%%%%%%%%%%%%%%%%%%%%%%%%%%%%%%%%%%%%%

In Fig.~\ref{fig:win3tev}, we plot the proton lifetime as functions of
the higgsino mass. Here, the wino, sfermion\footnote{
To be concrete, we regard $M_S$ as the stop mass, and all of the other
sfermion masses are assumed to be degenerate with $M_S$. Generally
speaking, stops are lighter than other sfermions, especially those of
the first and second generations. So, even though one relaxes the
degeneration assumption, one ends up obtaining a smaller proton decay
rate. }, and color-triplet Higgs
masses are set to be $M_2=3$~TeV, $M_S=10^3$~TeV, and $M_{H_C}=1.0\times
10^{16}~{\rm GeV}$, respectively. The solid lines correspond to $\tan
\beta=5,10,30$, and 50 from right-top to left-bottom,
respectively. Again the shaded region is excluded by the current
experimental bound, $\tau(p\to K^+ \bar{\nu})>4.0\times 10^{33}$~yrs
\cite{Abe:2011ts}. It is found that the lifetime considerably depends
on the mass of higgsino as well as the value of $\tan \beta$. It
illustrates that the higgsino contribution is dominant in a wide range
of parameter region. Indeed, the contribution gets more significant
as the higgsino mass is raised up. We also show a similar plot in the case of
$M_2=300$~GeV and $M_S=100$~TeV in Fig.~\ref{fig:win300gev}. In this
case, large $\tan\beta$ region is excluded even if the higgsino mass is
around 1~TeV, while with a rather small value of $\tan\beta$ the proton
lifetime easily exceeds the experimental limit. Anyway, we have found
that in the high-scale SUSY scenario, the minimal SUSY  SU(5) GUT is
still alive without any conspiracy of suppressing the dimension-five
operators.

%%%%%%%%%%%%%%%%%%%%%%%%%%%%%%%%%%
\section{Conclusions and Discussion}
\label{conclusion}
%%%%%%%%%%%%%%%%%%%%%%%%%%%%%%%%%%

In this work, we have evaluated the proton decay lifetime via the
dimension-five operators in the high-scale SUSY scenario. It is found
that the higgsino exchanging diagram gives rise to the dominant
contribution in a wide range of parameter region. After all, we have
revealed that the proton lifetime may evade the current experimental
limit and, thus, the minimal SUSY SU(5) GUT is not excluded in the high-scale
SUSY scenario.

In the $\mu_H \simeq M_S$ case, after $M_{H_C}$ being fixed, the proton
lifetime depends only on $M_S$ and $\tan \beta$. In fact, these two
parameters are also crucial for the prediction of the Higgs boson mass
in the high-scale SUSY scenario. Therefore, since now we know that the
mass of Higgs boson is 126~GeV, it is possible to relate the $M_S$ and
$\tan \beta$ in the present scenario \cite{Giudice:2011cg, Ibe:2011aa,
Ibe:2012hu, Ibanez:2013gf}. Further, if the mass spectrum is
somehow fixed, we are able to constraint $M_{H_C}$ by requiring the
gauge coupling unification {\cite{Hisano:2013cqa, Hisano:1992mh,
Hisano:1992jj}. Precise analyses in 
this direction enable us to predict proton decay rate in this scenario,
and future experiments may examine the prediction. Such kind of
model-dependent study is
carried out on another occasion.

While the dimension-five proton decay is suppressed by the heavy
sfermion masses, the dimension-six one through the $X$-boson exchange
does not suffer from such a suppression. As referred to above,
the GUT scale in the high-scale SUSY is slightly lower than the
ordinary one. Since the dimension-six proton decay lifetime scales as
$\propto M_X^4$ with $M_X$ the mass of $X$-boson, it may be
significantly enhanced even by a small change in the GUT scale. In such
a case, the $p\to \pi^0e^+$ mode may dominate the $p\to K^+\bar{\nu}$ mode,
and both of the modes might be searched in future experiments.
For instance, the expected sensitivities of the Hyper-Kamiokande
with ten years exposure \cite{Abe:2011ts} are $1.3\times 10^{35}$ and
$2.5\times 10^{34}$ years at 90~\% confidence level for the $p\to
e^+\pi^0$ and $p\to \bar{\nu}K^+$ modes,\footnote{
Recent improvements in the analysis of the $K^+\to \pi^+\pi^0$
decay channel may provide a better sensitivity for the $p\to
\bar{\nu}K^+$ mode: $3.2\times 10^{34}$ years at 90~\% confidence level \cite{MiuraBLV}.}
 respectively, which enable us to explore a wide range of
parameter region in high-scale SUSY models. 

After all, the minimal SUSY SU(5)  GUT is still quite promising, and the proton
decay experiments may reveal the existence of supersymmetry as well as
the grand unification.

~\\~\\
{\it Note Added:} While this work was being finalized, we noticed the
authors in Refs.~\cite{McKeen:2013dma, Liu:2013ula} discussed the
dimension-five proton decay in a similar context. In
Ref.~\cite{McKeen:2013dma}, they have just shown
dimensional analysis to constraint the dimension-five operators, while
in Ref.~\cite{Liu:2013ula}, the proton lifetime is examined in
supergravity unified models.

%%%%%%%%%%%%%%%%%%%%%%%%%%%%%%%%%%%%
\section*{Acknowledgments}
%%%%%%%%%%%%%%%%%%%%%%%%%%%%%%%%%%%%

We thank Norimi Yokozaki for stimulating us into this work.
N.N. is also grateful to Toru Goto for fruitful discussions.
The work of N.N. is supported by Research Fellowships of the Japan Society
for the Promotion of Science for Young Scientists. The work of
J.H. is supported by Grant-in-Aid for Scientific research from the
Ministry of Education, Science, Sports, and Culture (MEXT), Japan,
No. 20244037, No. 20540252, No. 22244021 and No. 23104011, and also by
World Premier International Research Center Initiative (WPI
Initiative), MEXT, Japan.

%%%%%%%%%%%%%%%%%%%%%%%%%%%%%%%%%%%%%%%%%%%%%%
\section*{Appendix}
\appendix
%%%%%%%%%%%%%%%%%%%%%%%%%%%%%%%%%%%%%%%%%%%%%

%%%%%%%%%%%%%%%%%%%%%%%%%%%%%%%%%%%%
\section{Renormalization Factors}
\label{Renormalization}
%%%%%%%%%%%%%%%%%%%%%%%%%%%%%%%%%%%%

Here we present the explicit expressions for the renormalization factors
defined in Eq.~\eqref{L6}. First, we write the renormalization factors
$A_R^{(i, j)}$ and $\overline{A}_R$ as the products of the long- and
short-distance renormalization factors:
\begin{align}
 A_R^{(i, j)}&\equiv A_L A_S^{(i, j)}~, \nonumber \\
 \overline{A}_R&\equiv \overline{A}_L \overline{A}_S~,
\end{align}
where $A_L$ and $\overline{A}_L$ represent the long-distance
QCD renormalization factors between the electroweak scale ($\mu=m_Z$
with $m_Z$ the $Z$-boson mass) and the scale of $\mu = 2$~GeV, while
$A_S^{(i, j)}$ and $\overline{A}_S$ correspond to the short-distance
renormalization effects between the electroweak and GUT scales. 
All of the effects are to be computed at one-loop level.

%%%%%%%%%%%%%%%%%%%%%%%%%%%%%%%%%%
\subsection{Long-range Factors}
%%%%%%%%%%%%%%%%%%%%%%%%%%%%%%%%%%%

First, we discuss the long-distance renormalization factors. The factors
consist of two effects; one is the running of the quark masses from $\mu
= 2$~GeV to $\mu= m_Z$, and the other is the renormalization effect of
the four-Fermi operators in Eq.~\eqref{L6} from $\mu = m_Z$ to $\mu
=2$~GeV. We neglect the QED corrections since the electromagnetic
coupling is much smaller than the strong coupling.

Before analyzing the renormalization effects, we first discuss the input
parameters for quark masses. In Ref.~\cite{PDG}, the values of the light
quark masses $m_q$~($q=u,d,s$) are given in the $\overline{\rm MS}$
scheme at $\mu\simeq 2$~GeV, while those for $c$- and $b$-quarks are
presented in the $\overline{\rm MS}$ at $\mu= m_c$ and $m_b$, 
respectively. Since we define $\overline{m}_q$ in Eq.~\eqref{L6} in the
$\overline{\rm DR}$ scheme, we convert the input parameters into those
in the $\overline{\rm DR}$ scheme. We use the one-loop relation
\cite{Yamada:1993uh}:
\begin{equation}
 \overline{m}_q(\mu)=m_q(\mu)\biggl(1-\frac{\alpha_s(\mu)}{3\pi}\biggr) ~,
\end{equation}
where $\alpha_s\equiv g_s^2/4\pi$ with $g_s$ the strong coupling
constant. For $c$- and $b$-quarks, we evolve the masses to $\mu=2$~GeV by
using the RGEs. For top quark, on the other hand, the pole mass is
displayed in Ref.~\cite{PDG}. The relation between the pole mass and the
$\overline{\rm DR}$ mass is given by \cite{Yamada:1993uh}
\begin{equation}
 m_t=\overline{m}_t (\mu)\biggl[
1+\frac{\alpha_s(\mu)}{3\pi}\biggl(
6\log\frac{\mu}{\overline{m}_t}+5
\biggr)
\biggr]~.
\end{equation}
Now we consider the QCD renormalization effects. 
The RGEs for the Wilson coefficients of the effective operators in
Eq.~\eqref{L6} at one-loop \cite{Abbott:1980zj} are given as\footnote{
Two-loop effects are also calculated in Ref.~\cite{Nihei:1994tx}, though
their contribution is found to be small.
} 
\begin{equation}
  \mu \frac{\partial}{\partial
  \mu}C=-
\frac{4g_s^2}{16\pi^2}C~.
\end{equation}
By using the equation, as well as the RGEs for the quark masses,  we
readily obtain the long-range renormalization factors $A_L$ and
$\overline{A}_L$:
\begin{align}
 A_L&=\frac{\overline{m}_{u_i}\overline{m}_{d_i}(m_Z)
\cdot C(2~{\rm  GeV})}
{\overline{m}_{u_i}\overline{m}_{d_i}(2~{\rm GeV})\cdot C(m_Z)}
=\biggl(\frac{\alpha_s(2~{\rm GeV})}{\alpha_s(m_b)}\biggr)
^{-\frac{18}{25}}
\biggl(\frac{\alpha_s(m_b)}{\alpha_s(m_Z)}\biggr)
^{-\frac{18}{23}}~,\nonumber \\
\nonumber \\
 \overline{A}_L
&=\frac{\overline{m}_{t}^2\overline{m}_{d_i}(m_Z)
\cdot C(2~{\rm  GeV})}
{\overline{m}_{t}^2\overline{m}_{d_i}(2~{\rm GeV})\cdot C(m_Z)}
=\biggl(\frac{\alpha_s(2~{\rm GeV})}{\alpha_s(m_b)}\biggr)
^{-\frac{6}{5}}
\biggl(\frac{\alpha_s(m_b)}{\alpha_s(m_Z)}\biggr)
^{-\frac{30}{23}}~.
\end{align}
Numerically, we have
\begin{equation}
 A_L=0.53,~~~~~~~
\overline{A}_L=0.34,
\end{equation}
at one-loop level.

%%%%%%%%%%%%%%%%%%%%%%%%%%%%%%%%%%%%
\subsection{Short-range Factors}
%%%%%%%%%%%%%%%%%%%%%%%%%%%%%%%%%%

Next, we evaluate the short-distance renormalization factors. They are
composed of three factors. First, the effective operators given at the
GUT scale receive the renormalization effects as they are taken down to
the electroweak scale. Second, the Yukawa couplings in the color-triplet Higgs exchanging process are determined through the running of
the couplings from the electroweak scale to the GUT scale. Third, the
interaction vertices in the one-loop diagrams in Fig.~\ref{fig:1loop}
are obtained by evolving the SU(2)$_L$ gauge coupling and the Yukawa
couplings according to the RGEs from the electroweak scale to the SUSY
breaking scale, $\mu= M_S$.

Let us begin with the running of the Yukawa couplings. Initial values
for the Yukawa coupling constants are given by 
\begin{align}
 y_{u_i}(m_Z)&=\frac{g_2}{\sqrt{2}m_W}\overline{m}_{u_i}(m_Z)~, \nonumber \\
 y_{d_i}(m_Z)&=\frac{g_2}{\sqrt{2}m_W}\overline{m}_{d_i}(m_Z)~,  \nonumber \\
 y_{e_i}(m_Z)&=\frac{g_2}{\sqrt{2}m_W}\overline{m}_{e_i}(m_Z)~.
\end{align}
In the SM, the Yukawa couplings flow according to the
following RGEs at one-loop level: 
\begin{align}
 \mu\frac{\partial}{\partial \mu}y_{u_i}&=\frac{1}{16\pi^2}y_{u_i}
\biggl[\frac{3}{2}(y_{u_i}^2 -y_{d_i}^2) +Y_2
 -\frac{17}{20}g^2_1-\frac{9}{4}g_2^2-8
 g_3^2\biggr] , \nonumber \\
 \mu\frac{\partial}{\partial \mu}y_{d_i}&=\frac{1}{16\pi^2}y_{d_i}
\biggl[\frac{3}{2}(y_{d_i}^2-y_{u_i}^2) +Y_2
 -\frac{1}{4}g^2_1-\frac{9}{4}g_2^2-8
 g_3^2\biggr] , \nonumber \\ 
 \mu\frac{\partial}{\partial \mu}y_{e_i}&=\frac{1}{16\pi^2}y_{e_i}
\biggl[\frac{3}{2}y^2_{e_i}+Y_2
 -\frac{9}{4}g^2_1-\frac{9}{4}g_2^2\biggr] ,
\label{yukawaSM}
\end{align}
where $Y_2$ is given as
\begin{equation}
 Y_2=\sum_{i}(3y_{u_i}^2+3y_{d_i}^2+y_{e_i}^2)~,
\end{equation}
and we neglect the off-diagonal components for simplicity. The equations
are also applicable when the renormalization scale exceeds the gaugino
masses. Above the higgsino mass, Eq.~\eqref{yukawaSM} is valid except
that $Y_2$ is modified to
\begin{equation}
  Y_2=\sum_{i}(3y_{u_i}^2+3y_{d_i}^2+y_{e_i}^2)
+\frac{3}{10}g_1^2+\frac{3}{2}g_2^2~.
\end{equation}
Here we assume the ordinary supersymmetric relation for the
gaugino-Higgs-higgsino couplings. The couplings may deviate the
relation when the SUSY breaking scale is much higher than the gaugino
and higgsino masses, but it is found that the deviation is usually not
so significant \cite{ArkaniHamed:2004fb, Giudice:2004tc}.

At the SUSY breaking scale, the Yukawa couplings $y_f$ are matched with
the supersymmetric ones, $\overline{y}_f$, as follows:
\begin{align}
 \overline{y}_{u_i}(M_S)&=\frac{1}{\sin\beta}{y}_{u_i}(M_S)~, \nonumber \\
  \overline{y}_{d_i}(M_S)&=\frac{1}{\cos\beta}{y}_{d_i}(M_S)~,  \nonumber \\
  \overline{y}_{e_i}(M_S)&=\frac{1}{\cos\beta}{y}_{e_i}(M_S)~.
\end{align}
Above $\mu =M_S$, the RGEs for the Yukawa couplings are given as
\begin{align}
 \mu\frac{\partial}{\partial \mu}\overline{y}_{u_i}&=\frac{1}{16\pi^2}
\overline{y}_{u_i}
\biggl[3
 \sum_{j}\overline{y}^2_{u_j}+3\overline{y}^2_{u_i}+\overline{y}^2
_{d_i} -\frac{13}{15}g^2_1-3g_2^2-\frac{16}{3}
 g_3^2\biggr] , \nonumber \\
 \mu\frac{\partial}{\partial \mu}\overline{y}_{d_i} &=\frac{1}{16\pi^2}
\overline{y}_{d_i}\biggl[\sum_{j}(3\overline{y}^2_{d_j}+\overline{y}^2_{e_j})
+ 3\overline{y}^2_{d_i}+\overline{y}^2_{u_i}
-\frac{7}{15}g^2_1-3g_2^2-\frac{16}{3}
 g_3^2\biggr] , \nonumber \\
 \mu\frac{\partial}{\partial \mu}\overline{y}_{e_i} &=\frac{1}{16\pi^2}
\overline{y}_{e_i}
\biggl[\sum_{j}(3\overline{y}^2_{d_j}+\overline{y}^2_{e_j}) 
+3 \overline{y}^2_{e_i} -\frac{9}{5}g^2_1-3g_2^2\biggr] .
\end{align}
Then, at the GUT scale, the Yukawa couplings $f_{u_i}$ and $f_{d_i}$ in
Eq.~\eqref{yukawadef} are defined by 
\begin{align}
 f_{u_i}&\equiv \overline{y}_{u_i}(M_{H_C})~,\nonumber \\
 f_{d_i}&\equiv \overline{y}_{d_i}(M_{H_C})~.
\end{align}

For the gauge couplings, the one-loop gauge coupling beta function
coefficients are given in the SM as
\begin{equation}
 b_a=(41/10, -19/6, -7)~,
\end{equation}
for U(1)$_Y$, SU(2)$_L$, and SU(3)$_C$, respectively. Above the gaugino threshold,
the coefficients are converted to\footnote{
When evaluating the renormalization factors, we take the gluino mass
equal to wino mass for simplicity.}
\begin{equation}
 b_a=(41/10, -11/6, -5)~,
\end{equation}
and after the higgsinos showing up, they lead to
\begin{equation}
 b_a=(9/2, -7/6, -5)~.
\end{equation}
Finally, in the MSSM, they are given as
\begin{equation}
b_a=(33/5, 1, -3)~.
\end{equation}

The short-distance renormalization factors for the four-Fermi operators
in Eq.~\eqref{L6} are presented in Ref.~\cite{Abbott:1980zj}. For
the effective operators
generated by the wino exchanging diagram, the renormalization factor
for the Wilson coefficient is 
\begin{equation}
 C(\mu)=\biggl(\frac{\alpha_3 (\mu)}{\alpha_3 (\mu_0)}\biggr)
^{-\frac{2}{b_3}}\biggl(\frac{\alpha_2 (\mu)}{\alpha_2 (\mu_0)}\biggr)
^{-\frac{15}{2b_2}}\biggl(\frac{\alpha_1 (\mu)}{\alpha_1 (\mu_0)}\biggr)
^{-\frac{1}{10b_1}}C(\mu_0)~,
\end{equation}
while for those induced by the higgsino exchange, we have
\begin{equation}
 \overline{C}(\mu)=\biggl(\frac{\alpha_3 (\mu)}{\alpha_3 (\mu_0)}\biggr)
^{-\frac{2}{b_3}}\biggl(\frac{\alpha_2 (\mu)}{\alpha_2 (\mu_0)}\biggr)
^{-\frac{9}{4b_2}}\biggl(\frac{\alpha_1 (\mu)}{\alpha_1 (\mu_0)}\biggr)
^{-\frac{11}{20b_1}}\overline{C}(\mu_0)~.
\end{equation}
In the case of $\mu > M_S$, the theory is to be regarded as
supersymmetric, and the renormalization factors are obtained as the
product of the wave function renormalizations of the fields in the
effective operators \cite{Grisaru:1979wc}. 
The wino contribution to the effective operators in Eq.~\eqref{L6} is
induced by the effective operators with a form like
\begin{equation}
  C_{ij} \int d^2\theta
\epsilon_{\alpha\beta\gamma}\epsilon_{rs}\epsilon_{tu}Q^{\alpha
  r}_iQ_i^{\beta s}Q_1^{rt}L^u_j~,
\end{equation}
with $i,j=2,3$, while the higgsino contribution is generated by
\begin{equation}
 \overline{C}\int d^2\theta
\epsilon^{\alpha\beta\gamma}\overline{U}_{3\alpha}
\overline{E}_3\overline{U}_{1\beta}\overline{D}_{l\gamma}~,
\end{equation}
with $l=1,2$. Then, the REGs for the Wilson coefficients of the
operators are
\begin{equation}
 \mu\frac{\partial}{\partial \mu}C_{ij}
=\frac{1}{16\pi^2}\bigl[
2(\overline{y}^2_{u_i}+\overline{y}_{d_i}^2)+\overline{y}_{e_j}^2
-8g_3^2-6g_2^2-\frac{2}{5}g_1^2
\bigr]C_{ij}~,
\end{equation}
and
\begin{equation}
 \mu\frac{\partial}{\partial \mu}\overline{C}
=\frac{1}{16\pi^2}\bigl[
2(\overline{y}^2_{t}+\overline{y}_{\tau}^2)
-8g_3^2-\frac{12}{5}g_1^2
\bigr]\overline{C}~.
\end{equation}
Note that in this case it is
important to take the Yukawa interactions into account since the
effective operators contain the third generation chiral
superfields. 

With the RGEs presented above, we finally compute the short-distance
renormalization factors as follows:
\begin{align}
 A_S^{(i,j)}&=\frac{m_W^2\sin 2\beta  f_{u_i}f_{d_j}}
{4\pi~\overline{m}_{u_i}\overline{m}_{d_j}(m_Z)}
\frac{\alpha_2(M_S)}{\alpha_2^2(m_Z)}
\cdot \frac{C(m_Z)}{C(M_S)}
\frac{C_{ij}(M_S)}{C_{ij}(M_{H_C})}~,\nonumber \\
\overline{A}_S&=\frac{m_W^4\sin^2 2\beta  f_{t}
\overline{y}_t(M_S)\overline{y}_\tau (M_S)f_{d_i}}
{(4\pi)^2\alpha_2^2(m_Z)\overline{m}_{t}^2\overline{m}_\tau
\overline{m}_{d_i}(m_Z)}
\cdot
\frac{\overline{C}(m_Z)}{\overline{C}(M_{H_C})}~.
\end{align}

%%%%%%%FIGURE%%%%%%%%%%%%%%%%%%%%%%
\begin{figure}[t]
\begin{minipage}{0.5\hsize}
\begin{center}
\includegraphics[height=75mm]{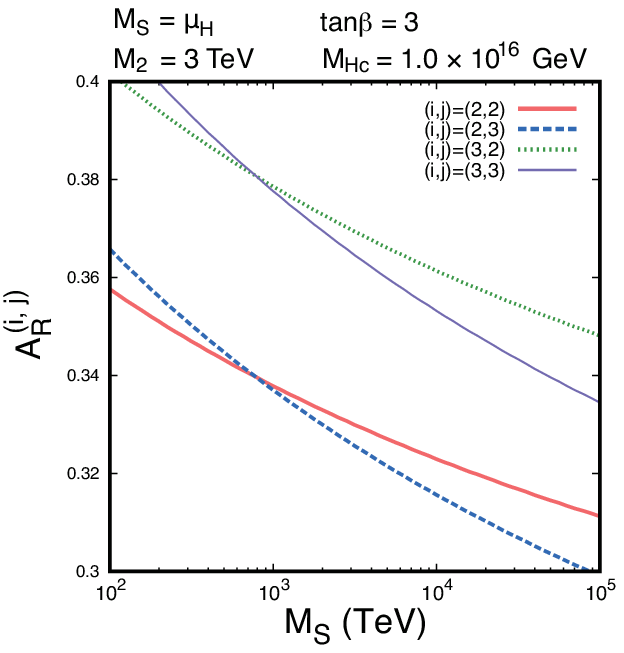}
\end{center}
\end{minipage}
\begin{minipage}{0.5\hsize}
\begin{center}
\includegraphics[height=75mm]{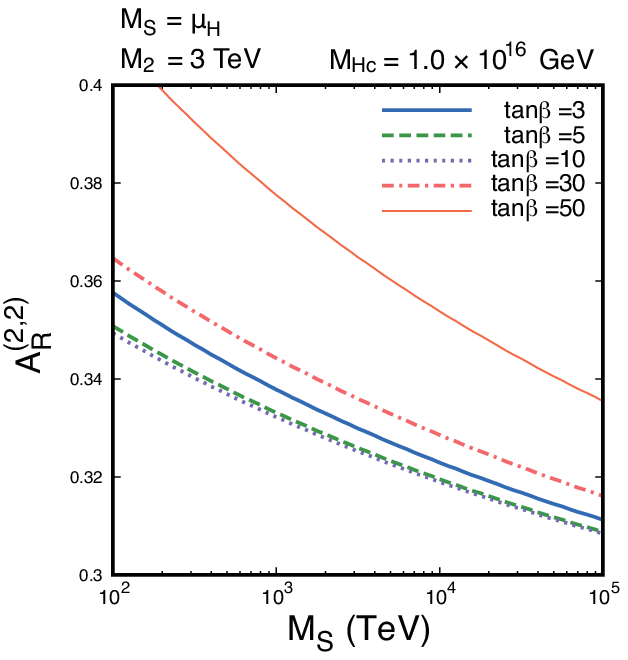}
\end{center}
\end{minipage}
\caption{Left:  $A_R^{(i,j)}$ with $(i,j)=(2,2),~(2,3),~(3,2),~(3,3)$ as functions of
$M_S$. Here we set $\tan \beta=3$, $M_S=\mu_H$, $M_2=3$~TeV, and $M_{H_C}=1.0\times
10^{16}$~GeV. Right: $A_R^{(2,2)}$ as functions of $M_S$. Same
 parameters as in the left graph are used except for $\tan \beta$. Each
 line corresponds to different values of $\tan \beta$ ($\tan\beta = 3,
 5, 10, 30,50$). }
\label{fig:ar}
\end{figure}
%%%%%%%%%%%%%%%%%%%%%%%%%%%%%%%%%%%%%%%%%%%%%%%%%%%%%%%%%%%%%%%

%%%%%%%FIGURE%%%%%%%%%%%%%%%%%%%%%%
\begin{figure}[t]
\begin{center}
\includegraphics[height=75mm]{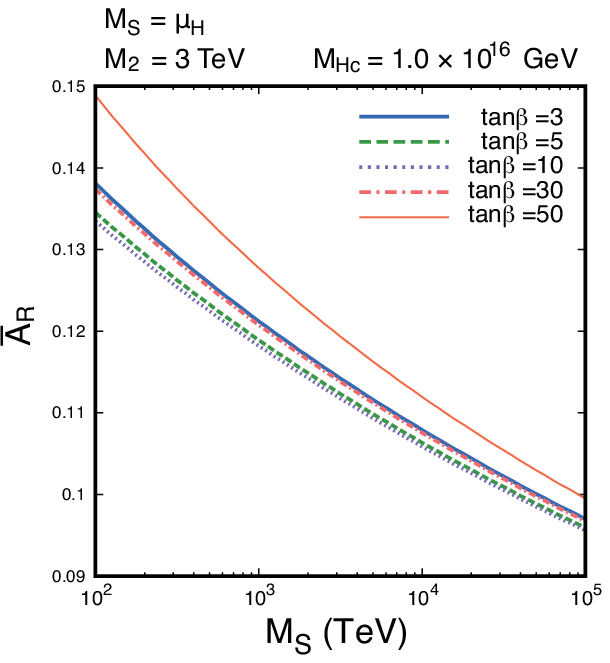}
\end{center}
\caption{$\overline{A}_R$ as functions of $M_S$. Here we set
 $M_S=\mu_H$, $M_2=3$~TeV, and $M_{H_C}=1.0\times
 10^{16}$~GeV. Each line shows different values of $\tan \beta$.}
\label{fig:arbar}
\end{figure}
%%%%%%%%%%%%%%%%%%%%%%%%%%%%%%%%%%%%%%%%%%%%%%%%%%%%%%%%%%%%%%%

With the results obtained above, we compute the renormalization
factors. We present $A_R^{(i,j)}$ and $\overline{A}_R$ in
Figs.~\ref{fig:ar} and \ref{fig:arbar}, respectively, as functions of
$M_S$. In both figures, we take $M_S=\mu_H$, $M_2=3$~TeV, and
$M_{H_C}=1.0\times 10^{16}$~GeV. In the left graph in Fig.~\ref{fig:ar},
each $A_R^{(i,j)}$ is presented with $\tan\beta$ fixed to be $\tan\beta
=3$, while in the right graph the behavior of $A_R^{(2,2)}$ is shown for
different values of $\tan \beta$ ($\tan\beta = 3, 5, 10,
30,50$). Similarly, each line in Fig.~\ref{fig:arbar} corresponds to
$\overline{A}_R$ evaluated with various $\tan \beta$'s. It is found that
the renormalization factors decrees as the SUSY scale increases, while
their dependence on $\tan \beta$ is somewhat complicated. In addition,
the left panel in Fig.~\ref{fig:ar} illustrates that the effects of the
third generation Yukawa couplings are significant.

%%%%%%%%%%%%%%%%%%%%%%%%%%%%%%%%%%%%%%%%%%%%%
\section{Formulae for Proton Decay Rate}
\label{Formulas}
%%%%%%%%%%%%%%%%%%%%%%%%%%%%%%%%%%%%%%%%%%%%

In this section, we display the formulae for the partial decay widths of
$p\to K^+ \bar{\nu}$ channels as well as the hadronic matrix elements
which we need to evaluate the decay widths. Let us start with the matrix
elements. We divide the derivation into two steps.\footnote{
Calculation of the matrix elements is also conducted by using the direct
method \cite{Aoki:2006ib}, in which the three-point correlation
functions relevant for proton decay are directly computed on the
lattice. Preliminary for the recent progress is reported in
Ref.~\cite{GUT2012}. 
} 
First, we express them in terms of the low-energy constants
$\alpha_p$ and $\beta_p$ by using the chiral perturbation techniques
\cite{Claudson:1981gh, Chadha:1983sj, Aoki:1999tw}:  
\begin{align}
 \langle K^+\vert \epsilon_{\alpha\beta\gamma}(u_L^\alpha d_L^\beta
)s^\gamma_L\vert p\rangle&=
\frac{\beta_p}{\sqrt{2}f_\pi}\biggl(
1+\frac{D+3F}{3}\frac{m_p}{M_B}
\biggr)P_L u_p~,
\nonumber \\
 \langle K^+\vert \epsilon_{\alpha\beta\gamma}(u_L^\alpha s_L^\beta
)d^\gamma_L\vert p\rangle&=
\frac{\beta_p}{\sqrt{2}f_\pi}\biggl(
\frac{2D}{3}\frac{m_p}{M_B}
\biggr)P_L u_p~,
\nonumber \\
 \langle K^+\vert \epsilon_{\alpha\beta\gamma}(u_R^\alpha d_R^\beta
)s^\gamma_L\vert p\rangle&=
\frac{\alpha_p}{\sqrt{2}f_\pi}\biggl(
1+\frac{D+3F}{3}\frac{m_p}{M_B}
\biggr)P_L u_p~,
\nonumber \\
 \langle K^+\vert \epsilon_{\alpha\beta\gamma}(u_R^\alpha s_R^\beta
)d^\gamma_L\vert p\rangle&=
\frac{\alpha_p}{\sqrt{2}f_\pi}\biggl(
\frac{2D}{3}\frac{m_p}{M_B}
\biggr)P_L u_p
~,
\end{align}
where $f_\pi\simeq 92.2$~MeV \cite{PDG} is the pion decay constant and
the baryon-meson couplings $D$ and $F$ are given as $D\simeq 0.80$ and
$F\simeq 0.47$, respectively. 
$m_p$ denotes the proton mass, while $M_B$ represents
the baryon mass parameter in the chiral Lagrangian, which we choose as 
$M_B\simeq (m_{\Sigma^0}+m_{\Lambda^0})/2$ with $m_{\Sigma^0}$ and
$m_{\Lambda^0}$ the masses of $\Sigma^0$ and $\Lambda^0$, respectively. 
$u_p$ is the four-component spinor wave function of proton, and $P_L$ is
the projection operator defined by $P_L\equiv (1-\gamma_5)/2$. The
low-energy constants $\alpha_p$ and $\beta_p$ are defined as
\begin{align}
 \langle 0\vert \epsilon_{\alpha\beta\gamma}(u^\alpha_R d^\beta_R)
u^\gamma_L \vert p\rangle&=\alpha_p P_L u_p~, \nonumber \\
 \langle 0\vert \epsilon_{\alpha\beta\gamma}(u^\alpha_L d^\beta_L)
u^\gamma_L \vert p\rangle&=\beta_p P_L u_p~, 
\end{align}
with $\vert 0\rangle$ the vacuum state. 
Second, we determine the constants $\alpha_p$ and $\beta_p$. We extract
them from the results of lattice simulations \cite{Aoki:2008ku}:
\begin{align}
 \alpha_p &=-0.0112\pm 0.0012_{(\rm stat)} \pm0.0022_{(\rm syst)}~{\rm
  GeV}^3~,\nonumber \\
 \beta_p &=0.0120\pm 0.0013_{(\rm stat)} \pm0.0023_{(\rm syst)}~{\rm
 GeV}^3~, 
\end{align}
where they are evaluated at $\mu = 2$~GeV.

With the matrix elements and the effective operators in
Eq.~\eqref{L6}, it is straightforward to derive the partial decay widths
of the $p\to K^+\bar{\nu}_\mu$ and $p\to K^+\bar{\nu}_\tau$ channels.
The result is
\begin{equation}
 \Gamma(p\to K^+\bar{\nu}_i)=
\frac{m_p \alpha_2^4\vert C_i\vert^2}{64\pi f_\pi^2 M^2_{H_C} m_W^4
\sin^22\beta} \biggl(
1-\frac{m_K^2}{m_p^2}
\biggr)^2~,
\end{equation}
with $(i=\mu, \tau)$ and
\begin{align}
 C_\mu&=2\beta_p F(M_2,M_S^2)\bigl\{
1+(D+F)\frac{m_p}{M_B}
\bigr\}\overline{m}_sV^*_{us}
\sum_{i=2,3}\overline{m}_{u_i}V_{u_id}V_{u_is}
e^{i\varphi_i}A^{(i,2)}_R~,
\label{cmu}
\end{align}
\begin{align}
 C_\tau&=2\beta_p F(M_2,M_S^2)\bigl\{
1+(D+F)\frac{m_p}{M_B}
\bigr\}\overline{m}_bV^*_{ub}
\sum_{i=2,3}\overline{m}_{u_i}V_{u_id}V_{u_is}
e^{i\varphi_{i}}A^{(i,3)}_R~\nonumber \\
&-\alpha_p\frac{\overline{m}_t^2\overline{m}_\tau V^*_{tb}e^{i\varphi_1}}
{m_W^2\sin 2\beta}F(\mu_H, M_S^2)\overline{A}_R
\biggl\{
\overline{m}_dV_{ud}V_{ts}\biggl(
1+\frac{D+3F}{3}\frac{m_p}{M_B}
\biggr)+\overline{m}_sV_{us}V_{td}\frac{2D}{3}\frac{m_p}{M_B}
\biggr\}~.
\label{ctau}
\end{align}

%%%%%%%%%%%%%%%%%%%%%%%%%%%%%%%%%%%%
{}
%%%%%%%%%%%%%%%%%%%%%%%%%%%%%%%%%%%%

\end{document}